\documentstyle[12pt]{article}
\newcommand{\beq}{\begin{equation}}
\newcommand{\eeq}{\end{equation}}
\newcommand{\bea}{\begin{eqnarray}}
\newcommand{\eea}{\end{eqnarray}}

\begin{document}
\title          {Continuum and Lattice Overlap for Chiral
                      Fermions on the Torus.}
%
\author{
\rule{0cm}{1.cm}
C.D. Fosco\\
{\normalsize\it
International Centre for  Theoretical Physics}\\
{\normalsize\it
P.O. Box 586, 34100 Trieste, Italy} }
\maketitle
\begin{abstract}
The overlap formulation is applied to calculate the chiral determinant
on a two-dimensional torus with twisted boundary conditions. We
evaluate first the continuum overlap, which is convergent and
well-defined, and yields the correct string theory result for both
the real and
imaginary parts of the effective action. We then show that the
lattice version of the overlap gives the continuum overlap results
in the limit when the lattice spacing tends to zero, and that
the subleading terms in that limit are irrelevant.
\end{abstract}
\newpage
\section{Introduction}
The overlap formalism~\cite{nar} is a  proposal, based on an
earlier idea of Kaplan~\cite{kap}, to define fermionic chiral
determinants.  When implemented on the lattice, it seems to
overcome the kinematical constraint stated by the Nielsen-Ninomiya
theorem~\cite{kn:Karsten.}, and thus could provide a suitable
framework to study non-perturbative phenomena in models
containing chiral fermions.

In this  method, the determinant of a chiral Dirac operator in
$2 d$ dimensions is defined as an overlap between the
Dirac vacuum states of two auxiliary Hamiltonians  acting on
Dirac fermions in $2 d + 1$ dimensions.

Different studies and tests have been performed on this new
formulation of the problem of evaluating chiral determinants,
both in its continuum and lattice versions.
In a Hamiltonian approach some tests for the continuum version
have been performed in $2$ and $4$ dimensions in references ~\cite{nn1}
and ~\cite{ran1}, respectively, showing that the continuum overlap
correctly reproduces some of the properties one expects for a
chiral determinant to have, namely chiral anomalies, zero modes, etc.
The $4$-dimensional continuum results have
also been confirmed  in a $5$-dimensional approach in ~\cite{chin}.
Recently, the relation between the phase of the continuum overlap
and the $\eta$-invariant has been elucidated~\cite{kaplan1}.
Regarding the lattice overlap, the main concern has been to show that
it does not suffer from any of the drawbacks  that afflicted previous
attempts to regulate fermions on the lattice. Analytic lattice
calculations have been carried out for slowly varying background
gauge fields~\cite{aoki,ran,ran2}. There
are also numerical calculations which confirm the overlap picture [1, 11].

In this paper  we present an expansion of a previous note~\cite{twist1}
on the case of fermions
with twisted boundary conditions on a torus.
Our exact results reinforce the conclusions of \cite{twi} where this problem
was studied numerically.
We consider here both the continuum and lattice version of the
overlap for twisted chiral fermions on the torus. We show that
the continuum overlap is already regularized and yields a finite
answer, that agrees with the one of more standard calculations~\cite{alv}.
The lattice regularization
is shown to lead to the continuum overlap in the continuum
limit. We also show that the subleading contributions to
the continuum calculation vanish when the lattice spacing tends
to $0$.

This paper is organized as follows: In section 2 some conventions and
definitions used in the formulation of the problem are presented.
The continuum overlap is dealt with in section 3, where the real
and imaginary parts of the effective action of the chiral Dirac
operator are calculated separately. In section 4 the lattice overlap is
considered, showing that it yields the  proper continuum limit for
both its real and imaginary parts. The subleading terms in the
small lattice spacing expansion are shown to be harmless.

\newpage
\section{Definitions and conventions.}
We summarize here some technicalities about the Dirac
equation and its twisted solutions on a torus with a flat metric.
The two-dimensional torus is coordinatized by two real
variables
\begin{equation}
\sigma^1 \;\;,\;\; \sigma^2 \;\;\;\; ,\;\;\;\; 0 \leq \sigma^{\mu} \leq 1 \;,
\label{1.1}
\end{equation}
and is equipped with the Euclidean metric
\begin{eqnarray}
ds^2 \,&=&\, \mid d \sigma^1 + \tau d \sigma^2 \mid^2 \,=\,
g_{\mu \nu} \; d\sigma^{\mu} d\sigma^{\nu}  \;,\nonumber\\
g_{\mu\nu} \,&=&\,
\left( \begin{array}{cc} 1 & \tau_1 \\ \tau_1 & \mid \tau \mid^2
\end{array} \right) \;\;\;\;\; g^{\mu\nu} \,=\,\frac{1}{\tau_2^2} \,
\left( \begin{array}{cc} \mid \tau \mid^2 & -\tau_1 \\ -\tau_1 & 1
\end{array} \right)
\label{1.2}
\end{eqnarray}
where $\tau = \tau_1 + i \tau_2$ and $\tau_2 > 0$. The associated
zweibeins $e^{\mu}_a$ then follow:
\begin{eqnarray}
g_{\mu \nu} \;e^{\mu}_a \, e^{\nu}_b \,&=&\, \delta_{a b} \;,\nonumber\\
e^{\mu}_1 \,=\, ( \,1\,,\, 0 \, ) \;& & \;
e^{\mu}_2 \,=\, (\,-\frac{\tau_1}{\tau_2} \,, \,\frac{1}{\tau_2} \,) \;\;.
\label{1.3}
\end{eqnarray}
To define the overlap, we need the vectorlike Dirac operator
\begin{equation}
\not \!\! D \,=\, \sigma^a \, e^{\mu}_a \, (\partial_{\mu} + i \,
A_{\mu}) \;,
\label{1.4}
\end{equation}
where the $\sigma^a$'s are the usual Pauli matrices
\begin{equation}
\sigma^1 \,=\,
\left( \begin{array}{cc}
0 & 1 \\ 1 & 0 \\ \end{array} \right) \;\; , \;\;
\sigma^2 \,=\,
\left( \begin{array}{cc}
0 & -i \\ i & 0 \\ \end{array}\right) \;\; , \;\;
\sigma^3 \,=\,
\left( \begin{array}{cc}
1 & 0 \\ 0 & - 1 \\ \end{array}\right) \;.
\label{1.5}
\end{equation}
We also define Dirac matrices consistent with the metric (\ref{1.2}), by
the equations
\begin{equation}
\gamma^{\mu} \, =\, \sigma^a \, e^{\mu}_a \;,\;\;\; \mu = 1,2 \;\;;\;\;\;
\gamma_5 \,=\, \sigma^3 \;.
\label{1.6}
\end{equation}
{}From (\ref{1.3}) and  (\ref{1.6}), some relations follow:
\begin{equation}
\{ \gamma^{\mu} , \gamma^{\nu} \} \;=\; 2 \, g^{\mu \nu} \;\;,\;\;
\{ \gamma^{\mu} , \gamma_5 \} \;=\; 0 \;,\;
{\rm tr} (\gamma^{\mu} \gamma^{\nu} \gamma_5) \,=\, \frac{2 i}{\tau_2} \,
\epsilon^{\mu \nu} \;,
\label{1.7}
\end{equation}
where $\epsilon^{\mu \nu}$ is the Levi-Civita symbol. Note that $\tau_2$
is the volume of the torus.

We would like to describe twisted fermions on the torus, i.e.,
the fermionic field has the boundary conditions
\begin{eqnarray}
\Psi ( \sigma^1 + 1 \,,\, \sigma^2 ) \,&=&\, - \, e^{2 \pi i
\varphi_1} \, \, \Psi (\sigma^1 \,,\, \sigma^2) \nonumber\\
\Psi ( \sigma^1 \,,\, \sigma^2 + 1) \,&=&\, - \, e^{2 \pi i
\varphi_2} \, \, \Psi (\sigma^1 \,,\, \sigma^2) \;\;,
\label{1.8}
\end{eqnarray}
where $\varphi_1$ and $\varphi_2$ are real numbers such that $0 \leq
\varphi_{\mu}
\leq 1 $. The `reference' boundary condition about which we define the
twistings is antiperiodicity in both directions, hence the minus signs
on the rhs in (\ref{1.8}). This is to assure that the `untwisted'
case ($\varphi_{\mu} \,=\, 0$), which will be used to normalize the twisted
determinant, does not have a zero mode.

To consider free fermions with the twistings (\ref{1.8}) is equivalent
(by a gauge transformation) to having fermions with antiperiodic
boundary conditions in the
background of a constant gauge field $A_{\mu} \;=\; 2 \pi \varphi_{\mu}$,
which is the description we shall adopt. Two
one-body Dirac Hamiltonians $H_{\pm} (\varphi)$ in a constant gauge field
$A_{\mu} \, = \, 2 \pi \varphi_{\mu}$ are then defined by
\begin{equation}
H_{\pm} (\varphi) \;=\; \gamma_5 \,  (\,\not \! \partial + 2 \pi i \not
\!\! \varphi \, \pm \, \mid \Lambda \mid \, ) \;,
\label{1.85}
\end{equation}
where $\Lambda$ is a constant with the dimensions of a mass.

Translational invariance of both the free and the constant-$A_{\mu}$
cases suggests the use of Fourier transforms.
The  Fourier transform of an antiperiodic function $f$, depending on
the coordinates $\sigma^{\mu}$ will be given by
\begin{equation}
f (\sigma) \;=\; \sum_n e^{i 2 \pi n_{\mu} \sigma^{\mu}} \, {\tilde f} (n)
\;\;,\;\;\;\;\; {\tilde f} (n) \;=\; \int d^2 \sigma \, e^{- i 2 \pi n_{\mu}
\sigma^{\mu}} \, f (\sigma) \;,
\label{1.9}
\end{equation}
where  $n^{\mu}$ runs over all the half-integers.

The free ($\varphi = 0$), positive ($u_{\pm}$), and negative ($v_{\pm}$)
eigenspinors in Fourier space satisfy
\begin{equation}
{\tilde H}_{\pm} (n) \, u_{\pm} (n) \,=\, \omega (n) \, u_{\pm} (n)
\;\; ; \;\;
{\tilde H}_{\pm} (n) \, v_{\pm} (n) \,=\, - \omega (n) \, v_{\pm} (n)
\label{1.10}
\end{equation}
where
\begin{eqnarray}
{\tilde H}_{\pm} (n) \, = \, \gamma_5 (2 \pi i \not \! n \pm \mid \Lambda \mid)
 \;& & \; \omega (n) \,=\, \sqrt{(2 \pi)^2 n^2  + \Lambda^2} \nonumber\\
u_{\pm}(n) = \frac{\omega (n) \pm | \Lambda | - 2 \pi i \not \! n}{
\sqrt{2 \omega (n) (\omega (n) \pm |\Lambda | )}} \, \chi & &
v_{\pm}(n) = \frac{\omega (n) \mp | \Lambda | + 2 \pi i \not \! n}{
\sqrt{2 \omega (n) (\omega (n) \mp | \Lambda | )}} \, \chi
\label{1.11}
\end{eqnarray}
and $\chi = \left(\begin{array}{c} 1\\0 \end{array} \right)$.

It should be evident from (\ref{1.8}) that
the eigenspinors of ${\tilde H}(n \mid \varphi)$, the Fourier transform
of $H_{\pm} (\varphi)$, are equal to those of $H_{\pm} (0)$ but with shifted
$n$'s: $n \to n + \varphi$,
\begin{eqnarray}
{\tilde H}_{\pm} (n\mid \varphi) \, u_{\pm} (n \mid \varphi) &=& \omega (n
|\varphi)
\, u_{\pm} (n\mid \varphi) \nonumber\\
{\tilde H}_{\pm} (n \mid \varphi) \, v_{\pm} (n \mid \varphi) &=& - \omega
(n \mid \varphi) \, v_{\pm} (n \mid \varphi) \nonumber\\
u_{\pm} (n \mid \varphi) \;=\; u_{\pm} (n + \varphi) \;&,&\;
v_{\pm} (n \mid \varphi) \;=\; v_{\pm} (n + \varphi)\;,
\label{1.12}
\end{eqnarray}
where $\omega (n\mid \varphi) = \omega (n + \varphi)$.

\section{Continuum overlap and effective action.}
Using the set of eigenspinors introduced in (\ref{1.12}),
we expand the fermionic fields
\begin{equation}
\Psi (\sigma) = \sum_n \left( b_{\pm} (n \mid \varphi) u_{\pm} (n \mid \varphi)
+ d_{\pm}^{\dagger} (n\mid \varphi) v_{\pm} (n\mid \varphi) \right)
e^{i 2 \pi n \cdot \sigma} \;,
\label{2.1}
\end{equation}
which contains the free ($\varphi = 0$) field expansion as
a particular case. The operators involved in (\ref{2.1}) satisfy the
anticommutation relations
\begin{eqnarray}
\{ \Psi (\sigma) \,,\, \Psi (\sigma') \} \;=\;0
\;& &\;
\{ \Psi (\sigma) \,,\, \Psi^{\dagger} (\sigma') \} \;=\; \delta (\sigma -
\sigma') \nonumber\\
\{ b_{\pm} (n \mid \varphi) \,,\, b_{\pm}^{\dagger} (n'\mid \varphi) \} \;=\;
\delta_{n,n'} \;& &\;
\{ d_{\pm} (n \mid \varphi) \,,\, d_{\pm}^{\dagger} (n'\mid \varphi) \} \;=\;
\delta_{n,n'} \nonumber\\
\{ b_{\pm} (n \mid \varphi) \,,\, b_{\pm} (n'\mid \varphi) \} \;=\; 0
\;& &\;
\{ d_{\pm} (k \mid \varphi) \,,\, d_{\pm} (n'\mid \varphi) \} \;=\; 0 \;,
\label{2.2}
\end{eqnarray}
where $\delta (\sigma - \sigma')$ is periodic
in $\sigma$, with period equal to $1$.
With these conventions, the corresponding Dirac vacua, i.e. the
vectors obtained by filling the negative energy states, are defined by
\begin{equation}
\mid \varphi \pm \rangle \;=\; \prod_n d_{\pm} (n \mid \varphi) \mid 0 \rangle
\;\;\;\;\;\; \mid 0 \pm \rangle \;=\; \prod_n d_{\pm} (n \mid 0) \mid 0 \rangle
\label{2.3}
\end{equation}
and the overlap definition of the normalized determinant becomes
\begin{equation}
\frac{\det D(\varphi)}{\det D(0)} \;=\; \lim_{\Lambda \to \infty}
\left[
\frac{\langle + \mid \varphi + \rangle}{\mid \langle + \mid \varphi + \rangle
\mid}
\;
\frac{\langle \varphi + \mid \varphi - \rangle}{\langle + \mid  - \rangle }
\;
\frac{\langle \varphi - \mid - \rangle}{\mid \langle \varphi - \mid - \rangle
\mid}
\right]
\label{2.4}
\end{equation}
where $D(\varphi)$ denotes the chiral Dirac operator:
$D(\varphi) \,=\, (\not \! \partial + 2 \pi i \not \! \varphi )\,
\displaystyle{\frac{(1 + \gamma_5)}{2}}$.

{}From (\ref{2.4}) we also define the Euclidean effective action
\begin{equation}
\Gamma (\varphi) \;=\; - \log \left[\frac{\det D(\varphi)}{\det D(0)} \right]
\;.
\label{2.5}
\end{equation}
Equations (\ref{2.4}) and (\ref{2.5}) can be written in terms of the
eigenspinors introduced in section 2. For example,\\ \\
$ \Gamma (\varphi,\Lambda) \,= $
\begin{equation}
- \sum_n \, \log \left(
\frac{v^{\dagger}_+ (n) v_+ (n+\varphi)}{\mid v^{\dagger}_+ (n) v_+(n+\varphi)
\mid}
\frac{v^{\dagger}_+ (n+\varphi) v_- (n+\varphi)}{v^{\dagger}_+ (n) v_-(n)}
\frac{v^{\dagger}_- (n+\varphi) v_- (n)}{\mid v^{\dagger}_- (n+\varphi)
v_-(n)\mid}
\right)
\label{2.6}
\end{equation}
where we have indicated the $\Lambda$-dependence of $\Gamma$ explicitly,
to make it
clear that the series on the rhs of (\ref{2.6}) shall be evaluated first,
taking the $\Lambda \to \infty$ limit afterwards:
\begin{equation}
\Gamma (\varphi) \;=\; \lim_{\Lambda \to \infty} \, \Gamma (\varphi , \Lambda)
\; .
\label{2.7}
\end{equation}
{}From (\ref{2.4}), (\ref{2.5}) and (\ref{2.6}), with the particular
phase conventions chosen for the eigenspinors,  the
real and imaginary parts of $\Gamma (\varphi , \Lambda)$ are
\begin{eqnarray}
{\rm Re}\, \Gamma (\varphi , \Lambda) &=& - \, \log \left(
\frac{\langle \varphi + \mid \varphi - \rangle}{\langle + \mid  -\rangle}
\right)
=\; - \sum_n \, \log \left( \frac{ v^{\dagger}_+ (n+\varphi) v_- (n +
\varphi)}{
v^{\dagger}_+ (n) v_- (n)} \right) \nonumber\\
{\rm Im} \Gamma (\varphi , \Lambda) &=& - {\rm Im} \, \log \left(
\frac{\langle + \mid \varphi + \rangle}{\mid \langle + \mid \varphi + \rangle
\mid} \;
\frac{\mid \langle - \mid \varphi - \rangle \mid}{\langle - \mid \varphi -
\rangle}
\right)
\nonumber\\ &=& -  {\rm Im} \, \sum_n \log \left(
\frac{v^{\dagger}_+ (n) v_+ (n+\varphi)}{\mid v^{\dagger}_+
(n) v_+ (n+\varphi)\mid} \frac{\mid v^{\dagger}_- (n) v_-
(n+\varphi)\mid}{v^{\dagger}_- (n) v_- (n+\varphi)} \right) \,.
\label{2.7}
\end{eqnarray}

We shall deal with the real and imaginary parts of the effective action
separatedly in the following subsections.

\subsection{Real part of the effective action.}
Using the explicit form for the eigenspinors
( (\ref{1.11}) and (\ref{1.12})) in (\ref{2.7}), we obtain
for ${\rm Re} \, \Gamma (\varphi , \Lambda)$ the following double-series
representation:
\begin{equation}
{\rm Re}\, \Gamma (\varphi , \Lambda) \;=\; \frac{1}{2} \,
\sum_n \, \log \left( 1 \,+\, \frac{ \Lambda^2 }{(2 \pi)^2 \, (n + \varphi)^2}
\right) \;
-\; \frac{1}{2} \,
\sum_n \, \log \left( 1 \,+\, \frac{ \Lambda^2 }{(2 \pi)^2  n^2}
\right) \; .
\label{2.8}
\end{equation}

With the explicit form of the metric tensor introduced in (\ref{1.2}),
and defining the parameter $\lambda \,=\, \displaystyle{\frac{\tau_2
\mid \Lambda \mid}{2 \pi}}$, we can rewrite the real part in a
more expanded form as\\ \\
$ {\rm Re} \Gamma (\varphi , \Lambda) \;= $
$$ \frac{1}{2}
\sum_{n_1,n_2} \, \log \left(
\frac{\mid \tau \mid^2  (n_1 + \varphi_1)^2 - 2 \tau_1 (n_1 + \varphi_1)(n_2 +
\varphi_2)
+(n_2 + \varphi_2)^2 + \lambda^2}{\mid \tau \mid^2  (n_1 + \varphi_1)^2 -
2 \tau_1 (n_1 + \varphi_1)(n_2 + \varphi_2)+(n_2 + \varphi_2)^2} \right)$$
\begin{equation}
\!\!\!\! - \frac{1}{2}
\sum_{n_1,n_2} \, \log \left(
\frac{\mid \tau \mid^2  n_1^2 - 2 \tau_1 n_1 n_2
+ n_2^2 + \lambda^2}{\mid \tau \mid^2  n_1^2 -
2 \tau_1 n_1  n_2 + n_2^2}\right) \;.
\label{2.9}
\end{equation}
Direct evaluation of the double series in (\ref{2.9}) is not a
straightforward task as it stands, but we can simplify the
calculation by taking derivatives with respect to $\lambda$ in
both sides of (\ref{2.9}), obtaining the differential equation
\begin{eqnarray}
\frac{\partial}{\partial \lambda} \, {\rm Re} \, (\,\varphi\,,\,\lambda\,)
\;&=&\; \lambda \, \left[ \sum_{n_1,n_2} \, \left(
\mid \tau (n_1 + \varphi_1) - (n_2 + \varphi_2) \mid^2 \,+\, \lambda^2
\right)^{-1}
\right. \nonumber\\
\;&-&\;\left. \sum_{n_1,n_2} \, \left(
\mid \tau n_1 - n_2 \mid^2 \,+\, \lambda^2 \right)^{-1}
\right]
\label{2.10}
\end{eqnarray}
where now the rhs of (\ref{2.10}) can indeed be evaluated by standard
methods. Let us first note that this expression is convergent, as
may for example be verified by expanding~\footnote{In our normalization
$0 \leq \varphi_\mu < 1$. Arguments based in series expansions in powers
of $\varphi_\mu$ will be used frequently in the following.} the first series
in powers of
$\varphi_{\mu}$. The term of order zero which might produce a logarithmic
divergence in the first series is exactly  cancelled by the
second series, and the successive terms produce series that behave
as $\mid n \mid^{-4}$ at worst, for large $\mid n \mid$ (in fact,
the term of order $q$ in $\varphi$ behaves like $\mid n \mid^{-2-q}$,
and only even $q$'s appear), thus they converge.

To evaluate the rhs of (\ref{2.10}) we shall first perform the summation
over one of the half-integer indices, $n_2$, say. The kind of expression
one needs to evaluate here will also appear in dealing with the imaginary
part of $\Gamma$. It has the following structure
\begin{equation}
f (x,y) \;=\; \sum_{n_2} \, \left[ \, (n_2 - x)^2 \,+\,y^2\,\right]^{-1}
\;,
\label{2.11}
\end{equation}
where $x$ and $y$ are functions of $n_1$ as well as of
$\varphi_{\mu}$ and $\lambda$. Then we resort to the procedure, familiar in
finite temperature quantum field theory, of converting the series
(\ref{2.11}) into an integral along a path in the complex plane,
by introducing a function which has poles at the half-integer real numbers:
\begin{equation}
f (x,y) \;=\; \oint_C \, \frac{d z}{2 \pi i}  \, \pi \, \tan (\pi z) \,
(z - x + i y)^{-1} \; (z - x - i y)^{-1} \;,
\label{2.12}
\end{equation}
where $C$ is a path that encircles all the poles of the $\tan$
counterclockwise~\footnote{This procedure is valid
provided $y > 0$, which is true for any $\Lambda \neq 0$.}, while
avoiding the poles at $z \,=\, x \pm i y$.
By deforming $C$, we can evaluate the integral by
knowing the residues at the poles $z_{\pm} \,=\, x \pm i y$. It yields
\begin{equation}
f (x,y) \;=\; \frac{\pi}{2 i \, y} \,
\left[ \; \tan (\,\pi (x + i y )\, )
\,-\, \tan (\,\pi (x - i y )\, ) \; \right] \;.
\label{2.13}
\end{equation}
Applying the result (\ref{2.13}) to (\ref{2.10}), we get \\ \\
$\frac{\partial}{\partial \lambda}{\rm Re} \,\Gamma (\varphi,\lambda)
\;=$
$${\displaystyle \frac{\pi\lambda}{2 i}}\;\times \sum_{n_1}
\left\{ \frac{1}{\eta (\varphi)}
\, \left[\;\;  \tan (\; \pi (\varphi_2 + \tau_1 (n_1 + \varphi_1) + i
\eta (\varphi) ) \;) \right. \right.$$
$$\left. -\tan (\;\pi (\varphi_2 + \tau_1 (n_1 + \varphi_1)
- i \eta (\varphi)
)\; )\;\;
\right] $$
\begin{equation}
\left.  -\,\frac{1}{\eta (0)}
\, \left[\;\; \tan (\;\pi (\tau_1 n_1  + i \eta (0))\;)
+\tan (\;\pi (\tau_1 n_1  - i \eta (0))\; )
\;\;\right] \right\} ,
\label{2.14}
\end{equation}
where $\eta (\varphi) \,=\, \sqrt{ \tau_2^2 (n_1 + \varphi_1)^2 + \lambda^2}$.
The next step is to integrate the differential equation (\ref{2.14}),
using the boundary condition ${\rm Re} \, \Gamma (\varphi , 0) \,=\,0$,
which follows from (\ref{2.9}).
This integration is elementary, the answer being
\begin{eqnarray}
{\rm Re} \, \Gamma (\varphi , \lambda) \,&=&\,
\frac{1}{2} \, \sum_{n_1} \, \left\{
\log \frac{\cos [ \pi ( \varphi_2 + \tau_1 (n_1 + \varphi_1) + i
\sqrt{ \lambda^2 + \tau_2^2 (n_1 +\varphi_1)^2} )]}{\cos [ \pi (\tau_1 n_1  + i
\sqrt{ \lambda^2 + \tau_2^2 n_1^2} )]} \right. \nonumber\\
&+& \,
\log \frac{\cos [ \pi ( \varphi_2 + \tau_1 (n_1 + \varphi_1) - i
\sqrt{ \lambda^2 + \tau_2^2 (n_1 +\varphi_1)^2} )]}{\cos [ \pi (\tau_1 n_1  - i
\sqrt{ \lambda^2 + \tau_2^2 n_1^2} )]} \nonumber\\
&-& \,
\log \frac{\cos [ \pi ( \varphi_2 + \tau_1 (n_1 + \varphi_1) + i
\tau_2 \mid n_1 +\varphi_1 \mid )]}{\cos [ \pi (\tau_1 n_1  + i
\tau_2 \mid n_1 \mid )]} \nonumber\\
&-& \,\left.
\log \frac{\cos [ \pi ( \varphi_2 + \tau_1 (n_1 + \varphi_1) - i
\tau_2 \mid n_1 +\varphi_1 \mid )]}{\cos [ \pi (\tau_1 n_1  - i
\tau_2 \mid n_1 \mid )]} \right\} \;\;.
\label{2.15}
\end{eqnarray}
So far no approximation has been made in deriving (\ref{2.15}), which is
exact for any $\lambda$ ($\propto |\Lambda|$). To obtain ${\rm Re}
\, \Gamma (\varphi)$ we shall
take the $\lambda \to \infty$ limit. Observing that (\ref{2.15})
may be decomposed into a $\lambda$-independent plus a $\lambda$-dependent
piece, we will find a more explicit expression for the former first, and
then take the limit $\lambda \to \infty$ for the latter\\ \\
$\left[ {\rm Re} \, \Gamma (\varphi , \lambda) \right]_{\lambda - indep}
\,=$
$$ -\frac{1}{2} \, \sum_{n_1} \,
\log \frac{\cos [ \pi ( \varphi_2 + \tau_1 (n_1 + \varphi_1) + i
\tau_2 \mid n_1 +\varphi_1 \mid )]}{\cos [ \pi (\tau_1 n_1  + i
\tau_2 \mid n_1 \mid )]}$$
\begin{equation}
-\frac{1}{2} \sum_{n_1}\, \log \frac{\cos [ \pi ( \varphi_2 + \tau_1
(n_1 + \varphi_1) - i \tau_2 \mid n_1 +\varphi_1 \mid )]}{\cos [ \pi (\tau_1
n_1  - i  \tau_2 \mid n_1 \mid )]} \,.
\label{2.16}
\end{equation}
Using the relation
\begin{equation}
\sum_{n} \, \log \frac{ \cos [ \pi ( \tau n + \alpha) ]}{\cos (\pi \tau n)}
\,=\, \log \frac{\vartheta (\alpha, \tau)}{\vartheta (0,\tau)} \;,
\label{2.17}
\end{equation}
where $n$ is half-integer and $\vartheta$ is a $\vartheta$-function (whose
infinite-product representation implies (\ref{2.17})) which
can be defined by the (single) series
\begin{equation}
\vartheta (\alpha , \tau) \;=\; \sum_n \, e^{ i \pi \tau n^2 \,+\,
2 \pi i n \alpha} \;,
\label{theta}
\end{equation}
one can verify after a bit of algebra that
\begin{equation}
\left[ {\rm Re} \, \Gamma (\varphi , \lambda) \right]_{\lambda - indep}
\;=\;- {\rm Re} \, \log \frac{\vartheta (\alpha, \tau)}{\vartheta (0,\tau)}
\;,\;\; \alpha \,=\, \tau \varphi_1 - \varphi_2 \;\;.
\label{2.18}
\end{equation}

Now we have to take the $\lambda \to \infty$ limit~\footnote{
If dimensions were introduced, this limit would
correspond to the ratio between the momentum spacing and  $\Lambda$
going to zero.} of the remaining,
$\lambda$-dependent part\\ \\
$\left[ {\rm Re} \, \Gamma (\varphi , \lambda) \right]_{\lambda - dep}
\;=$
$$ \frac{1}{2} \, \sum_{n_1} \,\left\{
\log \frac{\cos [ \pi ( \varphi_2 + \tau_1 (n_1 + \varphi_1) + i
\sqrt{ \lambda^2 + \tau_2^2 (n_1 +\varphi_1)^2} )]}{\cos [ \pi (\tau_1 n_1  + i
\sqrt{ \lambda^2 + \tau_2^2 n_1^2} )]} \right.$$
\begin{equation}
+\; \left.\,
\log \frac{\cos [ \pi ( \varphi_2 + \tau_1 (n_1 + \varphi_1) - i
\sqrt{ \lambda^2 + \tau_2^2 (n_1 +\varphi_1)^2} )]}{\cos [ \pi (\tau_1 n_1  - i
\sqrt{ \lambda^2 + \tau_2^2 n_1^2} )]} \right\} \;\;.
\label{2.19}
\end{equation}
The $|\lambda| \to \infty$ limit of (\ref{2.19}) is difficult to calculate
as it stands, because of the non-analytic behaviour in $\frac{1}{\lambda}$.
However, writing the $\cos$ in (\ref{2.19}) in terms of exponentials,
and keeping the leading terms (behaving as $e^{|\lambda|}$), one can prove
that only terms of up to order $2$ in $\varphi$ yield non-vanishing
contributions when $\lambda \, \to \infty$:
\begin{equation}
\left[ {\rm Re} \, \Gamma (\varphi , \lambda) \right]_{\lambda - dep}
\,=\, \pi \, \sum_{n_1} \,
\left[ 0 \, + \,
\frac{1}{2} \frac{\tau_2^2 \lambda^2}{(\lambda^2 +
\tau_2^2 n_1^2)^{\frac{3}{2}}} \, \varphi_1^2 \;+\; {\cal O}(\varphi^4)
\right]\;.
\label{2.20}
\end{equation}
The $\varphi$-independent term vanishes identically, whereas the
odd powers are absent by symmetry.  All the higher-order terms not shown in
(\ref{2.20}) are given by convergent series and vanish in the limit
$\lambda \to \infty$. Thus
\begin{equation}
\lim_{\lambda \to \infty}\left[ {\rm Re} \, \Gamma (\varphi , \lambda)
\right]_{\lambda - dep} \;=\; \varphi_1^2 \lim_{\lambda \to \infty}
\frac{\pi}{2} \, \sum_{n_1} \,
\frac{\tau_2^2 \lambda^2}{(\lambda^2 + \tau_2^2
n_1^2)^{\frac{3}{2}}} \,
\label{2.21}
\end{equation}
which can be calculated by replacing the series by an integral,
since the error of this replacement is made arbitrarily small
by just increasing $\lambda$. Then
\begin{equation}
\lim_{\lambda \to \infty}\left[ {\rm Re} \, \Gamma (\varphi , \lambda)
\right]_{\lambda - dep} \;=\; \varphi_1^2 \lim_{\lambda \to \infty}
\frac{\pi}{2} \, \int_{-\infty}^{+\infty} \, d x \,
\frac{\tau_2^2 \lambda^2}{(\lambda^2 + \tau_2^2
x^2)^{\frac{3}{2}}} \,=\, \pi\, \tau_2 \,\varphi_1^2 \;.
\label{2.22}
\end{equation}
Putting together the results for the $\lambda$-independent and
$\lambda$-dependent parts, we obtain the final result for the
real part of the effective action:
\begin{equation}
{\rm Re} \, \Gamma (\varphi , \lambda) \, = \, - \, {\rm Re} \,
\log \left( \, \frac{\vartheta (\alpha , \tau)}{\vartheta (0,\tau)}
\, \right) \; + \; \pi \tau_2 \varphi_1^2 \;.
\label{2.23}
\end{equation}

\subsection{Imaginary part of the effective action.}
In the continuum overlap definition of the imaginary part (\ref{2.7}),
we make use again of the explicit form of the eigenspinors to obtain
\begin{equation}
{\rm Im} \, \Gamma (\varphi , \Lambda) = {\rm Im} \, Z (\varphi , \Lambda)
\label{2.235}
\end{equation}
where\\ \\
$ Z (\varphi , \Lambda) \;= $
$$ \sum_n
\log \left[  (\omega (n) + |\Lambda |)(\omega (n+\varphi) + |\Lambda |)
\,+\, (2 \pi)^2 \, n \cdot (n + \varphi)
 +  \frac{(2 \pi)^2 i}{\tau_2}
\epsilon^{\mu \nu} n_{\mu} \varphi_{\nu} \right] $$
\begin{equation}
 - \sum_n
\log \left[  (\omega (n) - | \Lambda |)(\omega (n+\varphi) - |\Lambda |)
+ (2 \pi)^2 \, n \cdot (n + \varphi)
  + \frac{(2 \pi)^2 i}{\tau_2}
\epsilon^{\mu \nu} n_{\mu} \varphi_{\nu} \right]  \;.
\label{2.24}
\end{equation}

We now proceed to evaluate $Z (\varphi , \lambda)$, keeping in mind that
its real part is irrelevant to the calculation of the imaginary part
of $\Gamma$, and
we shall often ignore real terms in $Z$. This implies
that we shall not be concerned with the convergence of $Z$ in
(\ref{2.24}) (which looks badly divergent), but rather with the
convergence of its imaginary part only.
Again, it is not clear how to evaluate the series defining $Z$ as it
stands. However, a procedure similar to the one used in the calculation
of the real part can be applied, if one first simplifies expression
(\ref{2.24}). We cannot sum exactly the series for an arbitrary
$\Lambda$, as it was the case for the real part, due to the presence
of square roots. A simplification is achieved by taking the
$\Lambda \to \infty$ limit {\em before} summing over $n$. We are entitled
to interchange the order of summation and limit because the
imaginary part of $Z$ is convergent for any $\Lambda$,
even for $\Lambda \to \infty$. To show this, we first write the imaginary
part of $Z$ more explicitly:\\ \\
$ {\rm Im} \,Z (\varphi , \Lambda)$
$$=\; \sum_n \arctan \left[
\frac{(2 \pi)^2}{\tau_2} \, \frac{\epsilon^{\mu \nu}  n_\mu \varphi_\nu}{
(\omega (n) + \Lambda) (\omega (n+ \varphi) + \Lambda) \,+\,
(2 \pi)^2 n \cdot (n + \varphi)}\right] $$
\begin{equation}
- \sum_n \arctan \left[
\frac{(2 \pi)^2}{\tau_2} \, \frac{\epsilon^{\mu \nu}  n_\mu \varphi_\nu}{
(\omega (n) - \Lambda) (\omega (n+ \varphi) - \Lambda) \,+\,
(2 \pi)^2 n \cdot (n + \varphi)}\right]  \;.
\label{extra1}
\end{equation}
To study the convergence of each one of the series in (\ref{extra1})
(corresponding to the two possible signs of $\Lambda$),
we analyze the behaviour of the corresponding summands when
$n \to \infty$. For sufficiently large $n$, each one of the
expressions between brackets in (\ref{extra1}) behaves at
worst like $|n|^{-1}$, which can of course be made much
smaller than one. Thus in analyzing the convergence of the
series we can replace the $\arctan$'s by their small-arguments, i.e.,
\begin{eqnarray}
&\arctan& \left[\frac{(2 \pi)^2}{\tau_2}\, \frac{\epsilon^{\mu \nu}
n_\mu \varphi_\nu}{ (\omega (n) + \Lambda) (\omega (n+ \varphi) + \Lambda)
\,+\,
(2 \pi)^2 n \cdot (n + \varphi)}\right] \nonumber\\
&\simeq& \frac{(2 \pi)^2}{\tau_2} \, \frac{\epsilon^{\mu \nu}
n_\mu \varphi_\nu}{ (\omega (n) + \Lambda) (\omega (n+ \varphi) + \Lambda)
\,+\,
(2 \pi)^2 n \cdot (n + \varphi)}\;,
\label{extra2}
\end{eqnarray}
and analogously for the second term in (\ref{extra1}).
Now we study the convergence
of (the sum over $n$ of) (\ref{extra2}). Again we follow the approach
of performing an expansion in powers of $\varphi$, and showing that
the terms which might produce divergences actually vanish when
summed over $n$. It is
not difficult to realize that only even powers of $\varphi_\mu$
are allowed by symmetry, and that terms with $4$ powers or more
of $\varphi$ are convergent. Thus we only need to prove the convergence
of the second-order term
\begin{eqnarray}
\sum_n \, &\frac{(2 \pi)^2}{\tau_2}& \, \frac{\epsilon^{\mu \nu}
n_\mu \varphi_\nu}{ (\omega (n) + \Lambda) (\omega (n+ \varphi) + \Lambda)
\,+\,
(2 \pi)^2 n \cdot (n + \varphi)} \nonumber\\
&=& \frac{(2 \pi)^4}{\tau_2} \, \epsilon^{\mu \nu} \varphi_\nu \varphi^\lambda
\; \sum_n \, \frac{ 2 \,+\, \frac{\Lambda}{\omega (n)}}{(\omega (n)
+ \Lambda)^4} n_\mu n_\lambda \;+\; {\cal O} (\varphi^4) \;.
\label{extra3}
\end{eqnarray}
The quadratic term in (\ref{extra3}) vanishes by symmetry, because
\begin{equation}
\sum_n \, \frac{ 2 \,+\, \displaystyle{\frac{\Lambda}{\omega (n)}}}{(\omega (n)
+ \Lambda)^4} \; n_\mu n_\lambda \,\propto \,
g_{\mu \lambda} \;,
\label{extra4}
\end{equation}
and thus the second order term becomes proportional to
\begin{equation}
\epsilon^{\mu \nu} \varphi_\nu \varphi^\lambda g_{\mu \lambda}\;=\;
\epsilon^{\mu \nu} \varphi_\mu \varphi_\nu \;=\; 0 \;.
\label{extra5}
\end{equation}

Having shown that the series wich defines the imaginary part
of the effective action is convergent~\footnote{The large momentum
behaviour of the terms in the sum defining the imaginary part is much
milder than the ones corresponding to the real part. This may be
expected from the fact that the imaginary part is related to the
chiral anomaly.} for any $\Lambda$, we take
the limit $\Lambda \to \infty$ before summing over $n$. In this
limit the first log in (\ref{2.24})  yields a vanishing
contribution to the imaginary part of the effective action,
since
\begin{equation}
\omega (n) \; + \; |\Lambda | \;\; \to \;\; {\cal O} (|\Lambda |) \;\;
\label{ss1}
\end{equation}
when $|\Lambda | \; \to \; \infty $ and this suppress the imaginary
part of the first log by a power of $|\Lambda|^{-1}$. Regarding
the second log in (\ref{2.24}), the different sign in front
of $|\Lambda|$ produces the asymptotic behaviour
\begin{equation}
\omega (n) \; - \; |\Lambda | \;\; \to \;\; {\cal O} (|\Lambda |^{-1}) \;\;.
\label{ss2}
\end{equation}

Thus
when $\Lambda \to \infty$ the non-vanishing contributions originate
only from the second log in (\ref{2.24}) and they are given by the
imaginary part of~\footnote{We remind the reader that the divergence
of (\ref{2.25}) is irrelevant for our discussion as we are
interested in ${\rm Im} Z(\varphi , \Lambda)$, which as argued
above, is convergent. Similar arguments will be used again in the
next section.}
\begin{equation}
Z (\varphi , \Lambda) \, = \, - \sum_n \, \log \left[
n \cdot (n+ \varphi)
+ \frac{i}{\tau_2} \, \epsilon^{\mu \nu} n_{\mu} \varphi_{\nu} \right]
\;,
\label{2.25}
\end{equation}
Now we perform the summation over $n$. It is convenient to introduce
an auxiliary variable $\epsilon$ into (\ref{2.25}), in order
to be able to calculate this series:
\begin{equation}
Z (\varphi) \,=\, - \lim_{\epsilon \to 0} \, \sum_n \,
\log [ \epsilon^2 \, + \, n \cdot (n + \varphi) + \frac{i}{\tau_2}
\epsilon^{\mu \nu} n_{\mu} \varphi_{\nu} ] \;.
\label{2.26}
\end{equation}
The rest of the evaluation becomes akin to the one already performed
for the real part of the effective action. We differentiate
(\ref{2.26}) with respect to $\epsilon$:\\ \\
$\displaystyle{\frac{\partial}{\partial \epsilon}} Z (\varphi, \epsilon)$
$$=\; - 2 \epsilon \sum_n \, \left[ \epsilon^2 + \mid \tau \mid^2 n_1
(n_1 + \varphi_1) - \tau_1 n_1 (n_2 + \varphi_2)\right.$$
\begin{equation}
\left. - \tau_1 n_2 (n_1 + \varphi_1) +
n_2 (n_2 + \varphi_2) + i \tau_2 (n_1 \varphi_2 - n_2 \varphi_1)
\right]^{-1}
\label{2.27}
\end{equation}
and then perform the summation over the index $n_2$. The sum we
are faced with in (\ref{2.27}) is convergent for the same reasons
as the corresponding one for the real part.
The resulting
differential equation has however a different kind of boundary
condition, namely $\lim_{\epsilon \to \infty} {\rm Im}\, Z (\varphi ,
\epsilon ) = 0$.

After summing over $n_2$ and integrating between two values $\epsilon_2$,
$\epsilon_1$, we obtain
\begin{eqnarray}
Z (\varphi , \epsilon_1) \,-\, Z (\varphi , \epsilon_2) \;&=&\;
\sum_{n_1} \left[ - \log \cos \pi (\tau_1 n_1 +
\frac{\alpha}{2} + i s(\epsilon_1)) \right. \nonumber\\
&-& \left. \log \cos \pi (\tau_1 n_1 + \frac{\alpha}{2} - i
s(\epsilon_1)) \right]
\; - \; \left( \epsilon_1 \leftrightarrow \epsilon_2 \right)
\label{2.28}
\end{eqnarray}
where $s(\epsilon) \,=\, \sqrt{\epsilon^2 - (\frac{\alpha}{2} + i
\tau_2 n_1)^2}$.
Of course we are interested in the $\lim_{\epsilon \to 0} Z(\varphi ,
\epsilon)$,
so we derive it from (\ref{2.28}) by letting $\epsilon_1$ go to $0$
and $\epsilon_2$ to $\infty$:
\begin{eqnarray}
Z (\varphi) &=& \lim_{\epsilon \to 0} Z(\varphi , \epsilon)  \;=\;-\,\sum_{n_1}
\log \left[ \cos \pi (\tau n_1 + \alpha)\; \cos \pi {\bar \tau} n_1 \right]
\nonumber\\ &+& \lim_{\epsilon_2 \to \infty} \,
\sum_{n_1} \left[ \log \cos \pi (\tau_1 n_1 +
\frac{\alpha}{2} + i
\sqrt{\epsilon_2^2 - (\frac{\alpha}{2} + i \tau_2 n_1)^2}
) \right. \nonumber\\
&+& \left. \log \cos \pi (\tau_1 n_1 + \frac{\alpha}{2} - i
\sqrt{ \epsilon_2^2 - (\frac{\alpha}{2} + i \tau_2 n_1)^2}) \right] \;.
\label{2.29}
\end{eqnarray}
The contribution of the first sum in (\ref{2.29}) to the imaginary part
of $Z(\varphi , \Lambda)$ can be rewritten as
\begin{eqnarray}
- \, {\rm Im} \, \sum_{n_1} \, \log \left[
\cos \pi (\tau n_1 + \alpha)\; \cos \pi {\bar \tau} n_1\right]
&=& - \, {\rm Im} \, \sum_{n_1} \, \log [
\frac{\cos \pi (\tau n_1 + \alpha)}{\cos \pi \tau n_1} ]
\nonumber\\
&=& - {\rm Im} \, \log \left[ \frac{ \vartheta (\alpha , \tau) }{
\vartheta (0 , \tau)} \right] \;.
\label{2.30}
\end{eqnarray}
This fits exactly with the result for the $\lambda$-independent
part in (\ref{2.18}), to yield the
full $\vartheta$-function dependence of the effective action.
The limiting procedure for the second term in (\ref{2.29})
is analogous to the one followed in the calculation of
(\ref{2.19}-\ref{2.22}):
\begin{eqnarray}
\lim_{\epsilon_2 \to \infty} & & {\rm Im} \,
\sum_{n_1} \left[ \log \cos \pi (\tau_1 n_1 +
\frac{\alpha}{2} + i
\sqrt{\epsilon_2^2 - (\frac{\alpha}{2} + i \tau_2 n_1)^2}
) \right. \nonumber\\
&+& \left. \log \cos \pi (\tau_1 n_1 + \frac{\alpha}{2} - i
\sqrt{ \epsilon_2^2 - (\frac{\alpha}{2} + i \tau_2 n_1)^2}) \right] \nonumber\\
&=& 2 \pi \, {\rm Im} \, \lim_{\epsilon_2 \to \infty} \, \sum_{n_1} \,
\left[\sqrt{\epsilon_2^2 \,-\, (\frac{\alpha}{2} + i \tau_2 n_1)^2 } \,-\,
\sqrt{\epsilon_2^2 \,+\, \tau_2^2 n_1^2 } \right] \nonumber\\
&=& - {\rm Im} \, \frac{ \pi \alpha^2}{2 \tau_2} \nonumber\\
&=& - \, \tau_1 \pi \varphi_1^2 \, + \, \pi \varphi_1 \varphi_2 \;.
\label{2.31}
\end{eqnarray}
Substituting (\ref{2.30}) and (\ref{2.31}) in (\ref{2.29}), we get
\begin{equation}
{\rm Im} \, Z (\varphi) \,=\, - {\rm Im} \, \log
\frac{\vartheta (\alpha , \tau)}{\vartheta (0,\tau)} \,-\,\tau_1 \pi
\varphi_1^2 \,+\, \pi \varphi_1 \varphi_2 \,=\, {\rm Im} \, \Gamma (\varphi)
\;.
\label{2.315}
\end{equation}
The combination of this result with (\ref{2.23}) leads us to the
final form of the effective action
\begin{equation}
\Gamma (\varphi) \,=\, - \log \frac{\vartheta (\alpha , \tau)}{
\vartheta (0,\tau)} \,-\,i \pi \tau \varphi_1^2 \,+\, i \pi
\varphi_1 \varphi_2 \;.
\label{2.32}
\end{equation}
This agrees with the string theory evaluation of the determinant
of twisted fermions on a $2$-torus [9], and with the numerical
lattice calculation of ref.~\cite{twi}.

\newpage
\section{Lattice overlap.}
We now consider the problem of calculating the effective action when
the system is defined on a lattice of $N^2$ sites, which discretize
the torus. Although Equation (\ref{2.4}) remains formally the same
for this case, the Dirac vacua must be constructed from the
corresponding lattice eigenspinors instead. The discretized lattice
is defined by the set of points $\sigma^{\mu} \,=\, a \, t^{\mu}$,
where $t^{\mu}$ are integers such that $0 \leq t^{\mu} \leq N$, and
$a$ is the lattice spacing: $a \,=\, \frac{1}{N}$, where the last
relation follows from the requirement that for the continuum torus
$0 \leq \sigma^{\mu} \leq 1$. To the fermionic field
operators on the continuum, there correspond the lattice ones
$\Psi (\sigma) \; \to \; \Psi (a \, t)$, and the second-quantized
Dirac Hamiltonian is
\begin{equation}
{\hat H}_{\pm} (\varphi) \;=\; \sum_{t,s} \, \Psi^{\dagger} (a \, t) \,
H_{\pm}
(a \, (t - s) ) \, U(a t , a s) \, \Psi (a \, s) \; ,
\label{3.1}
\end{equation}
where $H_{\pm}$ is the lattice one-body Hamiltonian, which we
define in terms of its Fourier transform
\begin{eqnarray}
H_{\pm} (a \,(r-s)) &=& \frac{1}{N^2} \, \sum_{-N/2}^{+N/2} \,
{\tilde H}_{\pm} (n) \, e^{2 \pi i a n \cdot (r - s) } \nonumber\\
{\tilde H}_{\pm} (n) &=& \frac{1}{a} \, \sigma_3 \,
\left( i \not \!\! C (2 \pi a n) \,+\, B (2 \pi a n) \,\pm \,
a \mid \Lambda \mid \right) \;.
\label{3.2}
\end{eqnarray}
The link variables adopt a very simple, and moreover,
translation-invariant form
\begin{equation}
U (a r , a s ) \;=\;  e^{- i \int_{a s}^{a r} \, dx \cdot A} \;=\;
e^{- 2 \pi i a \varphi \cdot (r - s)} \;,
\label{3.3}
\end{equation}
and the functions $C_{\mu}$ and $B$ are defined by
\begin{eqnarray}
C_{\mu}(a n)  &=& \sin ( a n_{\mu} ) \nonumber\\
B (a n) &=& r \sum_{\mu = 1,2} \, (1 - \cos a n_{\mu} )
\label{3.4}
\end{eqnarray}
where $r$ is a number.
Our conventions for the 1-body Hamiltonian are so chosen as to assure
that it tends to its continuum version when $a \to 0$. The functions
$B$ and $C_\mu$ are dimensionless, the only dimensionful parameters
are $\Lambda$ and the lattice spacing. Note also that
the link variables may be introduced as part of the 1-body Hamiltonian
in Fourier space, just by computing the Fourier transform
\begin{eqnarray}
{\tilde H}_{\pm} (n \mid \varphi) &=& \sum_{t = 1}^N \, H_{\pm} (a t) \,
e^{- 2 \pi i a \varphi \cdot t} \; e^{- 2 \pi i a t \cdot n} \nonumber\\
&=& \sum_{t = 1}^N \, H_{\pm} (a t) \,
e^{- 2 \pi i a t \cdot (n + \varphi)} \,=\, {\tilde H}_{\pm} (n + \varphi)
\;,
\label{3.5}
\end{eqnarray}
which means that the lattice versions of relations (\ref{1.12}) hold.
In terms of
the eigenspinors of ${\tilde H}_{\pm} (n)$:
\begin{equation}
{\tilde H}_{\pm} (n) \, u_{\pm} (n) \,= \,\omega_{\pm} (n )
\, u_{\pm} (n) \;\;\;\;\;
{\tilde H}_{\pm} (n) \, v_{\pm} (n) \,=\, - \omega_{\pm}
(n) \, v_{\pm} (n) \;,
\label{3.6}
\end{equation}
where (see \cite{ran})
\begin{eqnarray}
\omega_{\pm} (n) \,&=&\, \frac{1}{a} \, \sqrt{ C^2 (a \,n) \,+\,
(B(a n) \,\pm \, a \, \mid \Lambda \mid )^2 } \nonumber\\
u_{\pm} (n) \,&=&\, \frac{\omega_{\pm} (n) \,+\, B(a \, n) \,\pm \,
a \mid \Lambda \mid \,-\, i \, \not \!  C (a \, n)}{
\sqrt{2 \omega_{\pm} (a \, n) \, (\omega_{\pm} (a \, n) \,+\,
B(a \, n) \, \pm \, a \mid \Lambda \mid )}} \; \chi \nonumber\\
v_{\pm} (n) \,&=&\, \frac{\omega_{\pm} (n) \,-\, B(a \, n) \,\mp \,
a \mid \Lambda \mid \,+\, i \, \not \!  C (a \, n)}{
\sqrt{2 \omega_{\pm} (a \, n) \, (\omega_{\pm} (a \, n) \,-\,
B(a \, n) \, \mp \, a \mid \Lambda \mid )}} \; \chi \,,
\label{3.7}
\end{eqnarray}
we define the field expansions
\begin{equation}
\Psi (r) \;=\; \frac{1}{N^2} \, \sum_{-\frac{N}{2}}^{+\frac{N}{2}} \,
\left[ b_{\pm} (n \mid \varphi) \, u_{\pm} (n \mid \varphi) \,+\,
d^{\dagger}_{\pm} (n \mid \varphi) \, v_{\pm} (n \mid \varphi) \right]
\, e^{2 \pi i a n \cdot r} \;.
\label{3.8}
\end{equation}
Now the relevant anticommutation relations read
\begin{eqnarray}
\{ \Psi (r) \, , \, \Psi (r') \} \,&=&\, 0  \nonumber\\
\{ \Psi (r) \, , \, \Psi^{\dagger} (r') \} &=& \delta (r - r')
\nonumber\\
\{ b_{\pm} (n \mid \varphi) \,,\, b^{\dagger}_{\pm} (n' \mid \varphi) \}
\,&=&\, \delta_{2 \pi} (n - n') \nonumber\\
\{ d_{\pm} (n \mid \varphi) \,,\, d^{\dagger}_{\pm} (n' \mid \varphi) \}
&=& \delta_{2 \pi} (n - n') \;,
\label{3.9}
\end{eqnarray}
where the periods of the $\delta$-functions are indicated
in their corresponding suffixes. The outcome of evaluating the
overlap with this conventions, and its $a \to 0$ limit are discussed
in the next two subsections, which deal with the real and imaginary
parts of the effective action separatedly.
\subsection{Real part of $\Gamma$.}
A straightforward calculation along the lines of the corresponding
continuum object yields
\begin{equation}
{\rm Re} \, \Gamma (\varphi , a) \;=\; - \, \sum_{-\frac{N}{2}}^{+
\frac{N}{2}} \, \log
\frac{v_+^{\dagger} (n + \varphi , a) v_- (n + \varphi , a)}{v_+^{\dagger}
(n , a) v_- (n , a)} \; ,
\label{3.10}
\end{equation}
where now we only display the dependence of $\Gamma$ on $\varphi$ and
the lattice spacing $a$, omitting the $\Lambda$-dependence since
this parameter will not be affected by the following treatment.
Using the explicit form of the lattice eigenspinors, we can
rewrite (\ref{3.10}) as \cite{ran}
\begin{equation}
{\rm Re} \, \Gamma (\varphi , a) \;=\; - \, \sum_{-\frac{N}{2}}^{+
\frac{N}{2}} \,\left\{  \log
\cos \beta (n + \varphi) \,-\, \log  \cos
\beta (n ) \right\} \;.
\label{3.11}
\end{equation}
where
\begin{eqnarray}
\cos \beta (n) &=& \sqrt{
\frac{\omega_+ (n) - a^{-1} B(n) - \mid \Lambda \mid}{2 \omega_+ (n)}
\frac{\omega_- (n) - a^{-1} B(n) + \mid \Lambda \mid}{2 \omega_- (n)}}
\nonumber\\
&+& \sqrt{
\frac{\omega_+ (n) + a^{-1} B(n) + \mid \Lambda \mid}{2 \omega_+ (n)}
\frac{\omega_- (n) + a^{-1} B(n) - \mid \Lambda \mid}{2 \omega_- (n)}}
\;.
\label{3.12}
\end{eqnarray}
Note that the factor $a^{-1}$ in front of $B$ is necessary to match
dimensions ($\Lambda$ is a mass).

We need to check that if we perform the summation over $n_\mu$
in (\ref{3.11}) and then take the limit $N \to \infty$, we
recover the continuum expression (22) for ${\rm Re} \, \Gamma (\varphi,
\Lambda)$. To this end first we note that the dominant
contributions to (\ref{3.11}) come from  the zeroes of
$\cos \beta (n)$ and $\cos \beta (n + \varphi)$. This happens
when $C_\mu$ and $B$ vanish simultaneously. But if
$r^2 >> \Lambda^2$, there will be no contribution from the doublers,
and the only zeroes of $\cos \beta$ will be at
$a n_\mu \to 0$. Hence we can expand (\ref{3.11}) in powers of
$a$. The leading, i.e., $a = 0$ term is convergent and coincides with
the continuum limit, since putting $a = 0$ everywhere
in (\ref{3.11}) amounts to replacing both $\omega_+$ and
$\omega_-$ by their common continuum counterpart $\omega (n)$,
and $B$ can be put to $0$ in ({\ref{3.12}), since its small-$a$
expansion begins with $a^2$.
With these replacements, the function which is summed in (\ref{3.11})
becomes identical to the one of the continuum overlap calculation
and moreover $N \,=\, a^{-1}$ tends to infinity, so that the
finite sum becomes a series.

One can show that the subleading terms
in $a$ are given by convergent series and hence make a vanishing
contribution when $a \to 0$. To see this we first note that
(\ref{3.11}) is an even function of $a$. Therefore, the first
subleading contribution will come from the coefficient of $a^2$
in the expansion of (\ref{3.11}) in powers of $a$.

The order $a^2$ term in the expansion  of the function
summed in (\ref{3.11}) in powers of $a$ turns out,
for large $n$, to behave like
\begin{eqnarray}
{\cal O} (a^2) &=& a^2 \; \frac{1}{2} \, \Lambda^2 \,
\left[ 2 r^2 (\, \frac{(n + \varphi)^4}{\omega^4 (n + \varphi)}
- \frac{n^4}{\omega^4 (n)}\, )
\,+ \, \frac{n_\mu (n^\mu)^3 + (n_\mu)^3 n^\mu}{n^2 \omega^2(n)}
\right. \nonumber\\
&-& \left.
\frac{ (n + \varphi)_\mu ((n + \varphi)^\mu)^3 + ((n + \varphi)^3)_\mu
(n + \varphi)^\mu}{(n +\varphi)^2 \omega^2 (n + \varphi)}
\right] \;.
\label{ex1}
\end{eqnarray}

Then by expanding (\ref{ex1}) in powers of
$\varphi$, the order $0$ term in $\varphi$ is zero, since it is
killed by the normalization we use for the determinant.
The term linear in $\varphi$ vanishes when summed over
$n$, as can be seen also from the fact that (\ref{3.11})
is invariant under $\varphi \to -\varphi$. But then the term
quadratic in $\varphi$ carries a behaviour $n^{-2}$ for large
$n$, and thus the sum over $n$ is at worst logarithmically
divergent with $1/a$. When $a$ tends to zero this logarithmic
factor is killed by the $a^2$ power multiplying this term.
Other powers of
$\varphi$ tend also to zero, since the term of order $q$ in
$\varphi$ carries a $1/n^q$ large-$n$ behaviour.
\subsection{Imaginary part of $\Gamma$.}
Using the explicit form of the lattice eigenspinors, we
can write the imaginary part of the effective action
for finite lattice spacing as follows
\begin{equation}
{\rm Im} \, \Gamma (\varphi , \Lambda , a) \;=\; {\rm Im} \,
\sum_{-\frac{N}{2}}^{+\frac{N}{2}} \,
\left[ \log G (n,\varphi,\Lambda,a)
\,-\, \log G (n,\varphi,-\Lambda,a) \right]
\;.
\label{3.13}
\end{equation}
where
\newpage
$ G (n,\varphi,\Lambda,a) \; = $
$$ [\, \omega_+ (n) +  \Lambda -
a^{-1} B(n) \,]
[\, \omega_+ (n+\varphi) +   \Lambda -
a^{-1} B(n+\varphi)\, ] $$
\begin{equation}
+ a^{-2} C(2 \pi a n) \cdot  C(2 \pi a (n+\varphi) )
\,+\,
\frac{i \epsilon^{\mu \nu} }{a^2 \tau_2}
\, C_{\mu} (2 \pi a n) \, C_{\nu} ( 2 \pi a (n + \varphi))
\label{3.14}
\end{equation}
It is immediate to realize that, again for the imaginary part, the
limit $a \to 0$ yields the continuum overlap definition which
was already evaluated. It is only necessary to realize that both
$\omega_+$ and $\omega_-$ tend to their common continuum limit, and that the
contributions from $B$ are suppressed by an extra power of $a$. So let us
discuss the subleading terms.
For any $a \neq 0$, one sees that
\begin{equation}
G ( n , \varphi , - \Lambda , a) \; = \;
G ( n , \varphi ,  \Lambda , - a) \;,
\label{3.15}
\end{equation}
therefore for $a \neq 0$ (\ref{3.15}) can be used in (\ref{3.14})
and (\ref{3.13}) to deduce:
\begin{equation}
{\rm Im} \, \Gamma (\varphi , \Lambda, -a) \; = \; -
{\rm Im} \, \Gamma (\varphi , \Lambda, a) \; ,
\end{equation}
thus only {\em odd} powers of $a$ (except for the $a = 0$-term)
may appear in a small-$a$ series. The first
subleading term might then be of order $a$. A calculation of
this order-$a$ term for (\ref{3.14}) yields a function with
the structure
\begin{equation}
{\cal O} (a) \;=\; a \, \sum_{n_1,n_2} \, f (n,\Lambda, n+ \varphi ) \,
\epsilon^{\mu \nu} n_{\mu} \varphi_{\nu} \;,
\label{3.16}
\end{equation}
where $f$ is given by:\\
$f (n, \Lambda, n+\varphi) = \frac{r}{\tau_2} \,
( \frac{n^2}{\omega (n)} + \frac{(n +\varphi)^2}{\omega (n+ \varphi)} )
\times$
$$\left[
\frac{ ( \omega (n + \varphi) + \Lambda ) ( \omega (n ) + \Lambda )}{
[( \omega (n + \varphi) + \Lambda ) ( \omega (n ) + \Lambda ) + n
\cdot (n + \varphi) ]^2 + n^2 \varphi^2 - (n \cdot \varphi)^2}
\right.$$
\begin{equation}
- \;\left.
\frac{ ( \omega (n + \varphi) - \Lambda ) ( \omega (n ) - \Lambda )}{
[( \omega (n + \varphi) - \Lambda ) ( \omega (n ) - \Lambda ) + n
\cdot (n + \varphi) ]^2 + n^2 \varphi^2 - (n \cdot \varphi)^2}
\right]
\;.
\label{exx}
\end{equation}

Equation (\ref{exx}) is  quite a  cumbersome scalar function of its arguments,
but we only need to know that it behaves as $1/n$ for large
$n$, and it depends on $\varphi$ only through the combination $n + \varphi$.
Thus the series (\ref{3.16}) appears to be quadratically divergent. Note
that the imaginary part of (\ref{3.14}) is an even function of $\varphi_{\mu}$,
so, even though the terms of order two and higher in an expansion of
$f$ in powers of $\varphi$ will tend to zero as for the real part case,
we are still left with the term of order one. This will however
vanish, because expanding $f$ to first order in $\varphi$ will yield
for (\ref{3.16})
\begin{equation}
{\cal O} (a) \;\sim\; a \, \sum_{n_1,n_2} \, h (n,\Lambda) \,
\epsilon^{\mu \nu} n_{\mu} \varphi_{\nu} n \cdot \varphi \;,
\label{3.17}
\end{equation}
where $h$ is also a scalar function but of $n$ and $\Lambda$, whose
form is not relevant to our argument. Since one can prove that
\begin{equation}
\sum_{n_1,n_2} h(n,\Lambda)\,  n^{\mu} n^{\nu} \;\propto \; g^{\mu \nu}
\;,
\label{3.18}
\end{equation}
where $g$ is the metric tensor. Inserting (\ref{3.18}) into
(\ref{3.17}) yields $0$ for this term, since
$\epsilon^{\mu \nu} \varphi_{\mu} \varphi_{\nu} \,=\,0$.

This completes the proof of the vanishing of the subleading term
for the imaginary part.
\section*{Acknowledgements.}
I am grateful to S. Randjbar-Daemi for many useful discussions
and suggestions, and to
H. Neuberger and R. Narayanan for reading the manuscript,
and useful correspondence.

\end{document}